\documentclass{PoS}

\title{DPHEP: From Study Group to Collaboration}
\ShortTitle{DPHEP: From Study Group to Collaboration}

\author{\speaker{David M. South}\thanks{on behalf of the DPHEP Collaboration.}\\
  Deutsches Elektronen Synchrotron, Notkestrasse 85, 22607 Hamburg, Germany\\
  E-mail: \email{david.south@desy.de}}

\abstract{The international study group on data preservation in high energy physics, DPHEP, achieved a milestone
    in 2012 with the publication of its eagerly anticipated large scale report, which contains a description of
    data preservation activities from all major high energy physics collider-based experiments and laboratories.
    DPHEP will evolve to a new collaboration structure in 2013. The formation of the study group is described, as well
    as some of the key messages from the report focussing on the physics case for the preservation of high
    energy physics data and a description of the different preservation models. Finally, the future working directions
    of the new collaboration are outlined.}

\FullConference{The European Physical Society Conference on High Energy Physics -EPS-HEP2013\\
		18-24 July 2013\\
		Stockholm, Sweden}

\begin{document}

\section{Introduction} 
\label{sec:intro}

The problem of data persistence and preservation is not new, but is becoming more prominent 
with the advent of so called big data, in particular within the applied sciences.
However, until recently high energy physics (HEP) had little or no tradition or clear model of
long term preservation of data in a meaningful and useful way, and the data from the majority
of older experiments have simply been lost.
The preservation and long term access of HEP data, has so far not been part of the planning, software
design or budget of particle physics experiments and such initiatives have in the main not been undertaken
by the original collaboration as a whole, but rather by a few individuals after the end of data taking and
with varying degrees of success.
This is despite several clear scenarios where preservation of HEP data is beneficial for a number of reasons.
After many decades of neglect with respect to other scientific disciplines, data preservation is now a 
rapidly emerging field in HEP, where the DPHEP Study Group~\cite{dphep} is now established as the coherent
multi-laboratory, multi-experiment body to examine this issue. 
In 2013 DPHEP is making the transition to a new collaboration structure.
These proceedings briefly describe the formation of the study group, some key messages from the group's
report on the current status of data preservation in HEP, and the future working directions of the new
collaboration.

\section{The DPHEP Study Group} 
\label{sec:dphepstudygroup}

The start of the 21st century saw the end of operation of several particle colliders including
LEP~($e^{+}e^{-}$, data taking ended in 2000), HERA~($e^{\pm}p$, 2007), PEP-II~($e^{+}e^{-}$, 2008),
KEKB~($e^{+}e^{-}$, 2010) and the Tevatron~($p\bar{p}$, 2011), providing unique data sets in
terms of initial state particles or centre of mass energy or both.
As the experiments at each of these colliders continued to publish their final results and conclude their
core physics programmes, the question of what should be done with the data naturally presented itself.
Inspired by a lack of concrete solutions or guidelines to the problem of data preservation in HEP,
an international study group on data preservation and long term analysis in high energy physics,
DPHEP~\cite{dphep}, was formed at the end of 2008 to address the issue in a systematic way.
%
%
The composition of the group was initially driven by BaBar and the HERA experiments H1, ZEUS and
HERMES, who were soon joined by Belle, BES-III and the Tevatron experiments CDF and D{\O}.
The LEP experiments are also represented in DPHEP and the LHC experiments ALICE, ATLAS, CMS
and LHCb joined the study group in 2011.
The laboratories and associated computing centres at BNL, CERN, DESY, Fermilab, JLAB, KEK and SLAC
are all also members of DPHEP, in addition to several funding agencies. 
A series of seven workshops have taken place since 2009 and DPHEP is officially endorsed with a
mandate by the International Committee for Future Accelerators, ICFA.
%
%
The initial findings of the study group were summarised in a short interim report in
December 2009~\cite{dpheppub1} and a full status report was released in
May 2012~\cite{dpheppub2}.
The full report contains: a tour of data preservation activities in other fields; an expanded
description of the physics case; a guide to defining and establishing data preservation
principles; updates from the experiments and joint projects, as well as person-power
estimates for these and future projects; the proposed next steps to fully establish
DPHEP in the field.
The physics case for data preservation and alternative preservation models are briefly
described in the following sections; further details can be found in the 2012 DPHEP
report~\cite{dpheppub2}.

\section{Building the physics case for data preservation}
\label{sec:physicscase}

The main motivation behind this project, and indeed any HEP data preservation initiative, is the possibility
of new physics results.
There are several scenarios where the preservation of experimental HEP data would be advantageous
to the particle physics community.

Data preservation is beneficial to the long term completion and extension of the physics programme
of an experiment.
In the case of the LEP experiments a considerable tail exists in the publication rate, which continues today and a
similar trend is now predicted by, among others, BaBar and the HERA experiments.
In particular, it is typical that precision analyses continue long after the end of data taking,
in order to make use of the full statistical power and the best knowledge of systematic uncertainties.
Up to $10$\% of papers are finalised in the post-collisions period, and prolonging the availability of the data
may result in a gain in scientific output of an experiment.

\begin{figure}[h]
  \centering
    \includegraphics[width=0.43\textwidth]{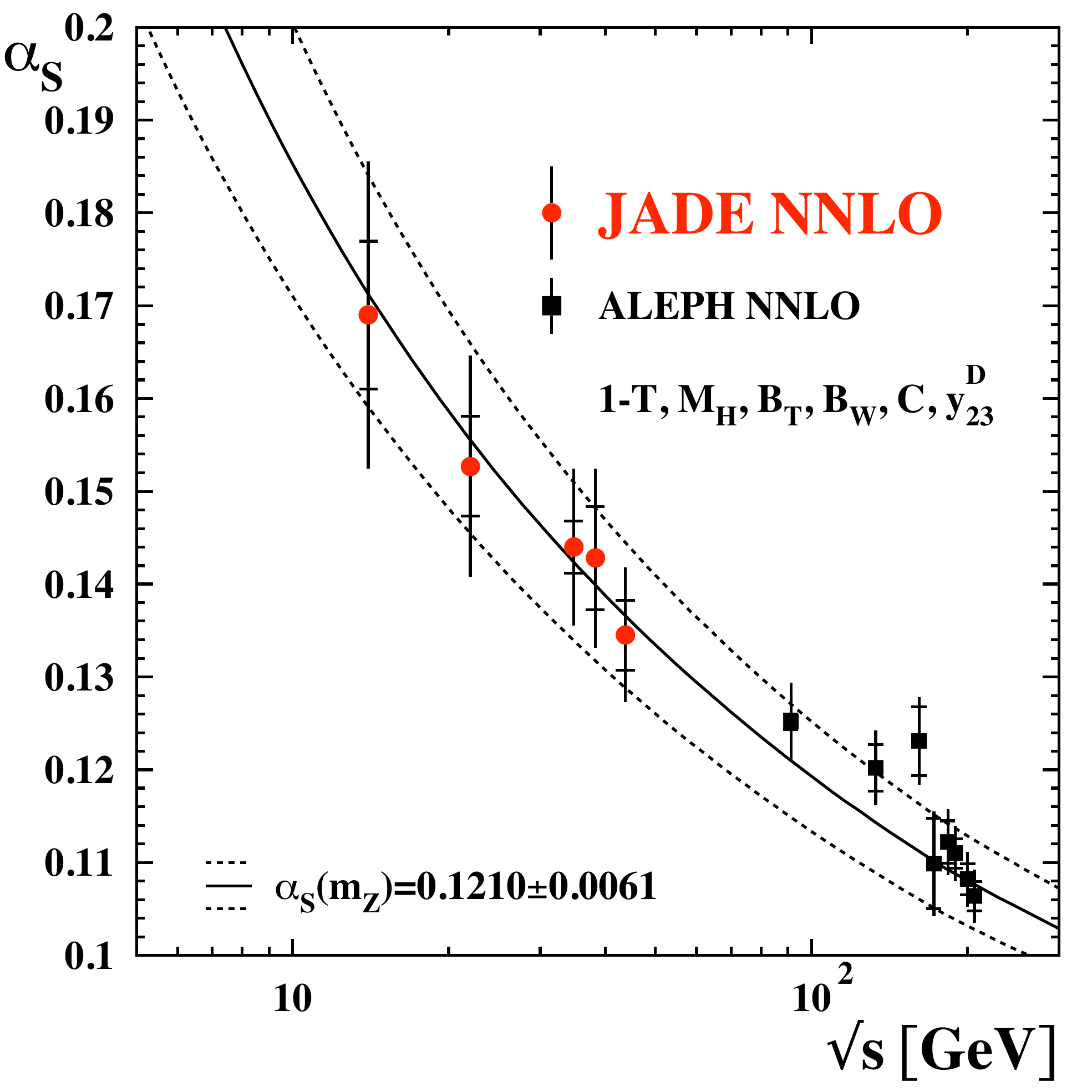}
    \includegraphics[width=0.41\textwidth]{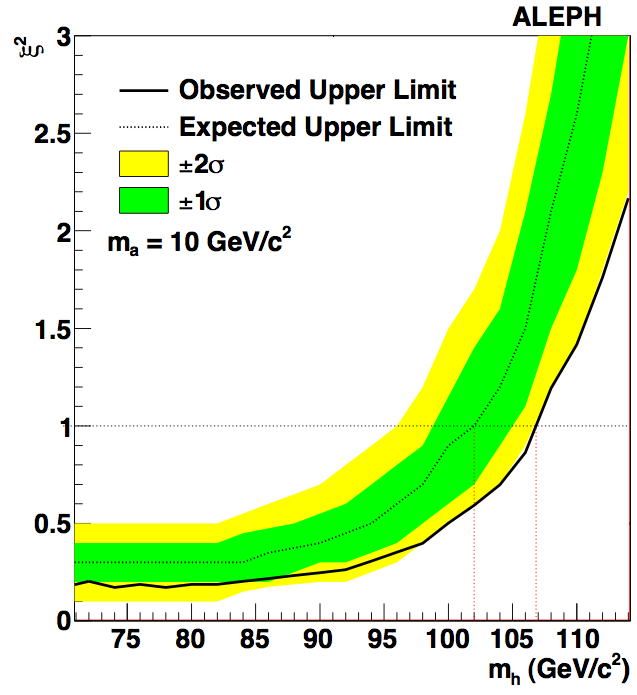}
    \includegraphics[width=0.41\textwidth]{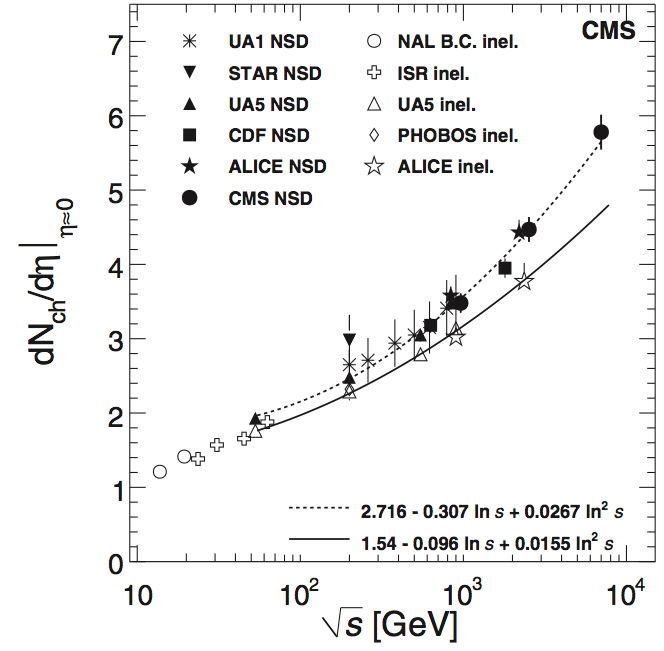}
    \hspace{0.2cm}
    \includegraphics[width=0.415\textwidth]{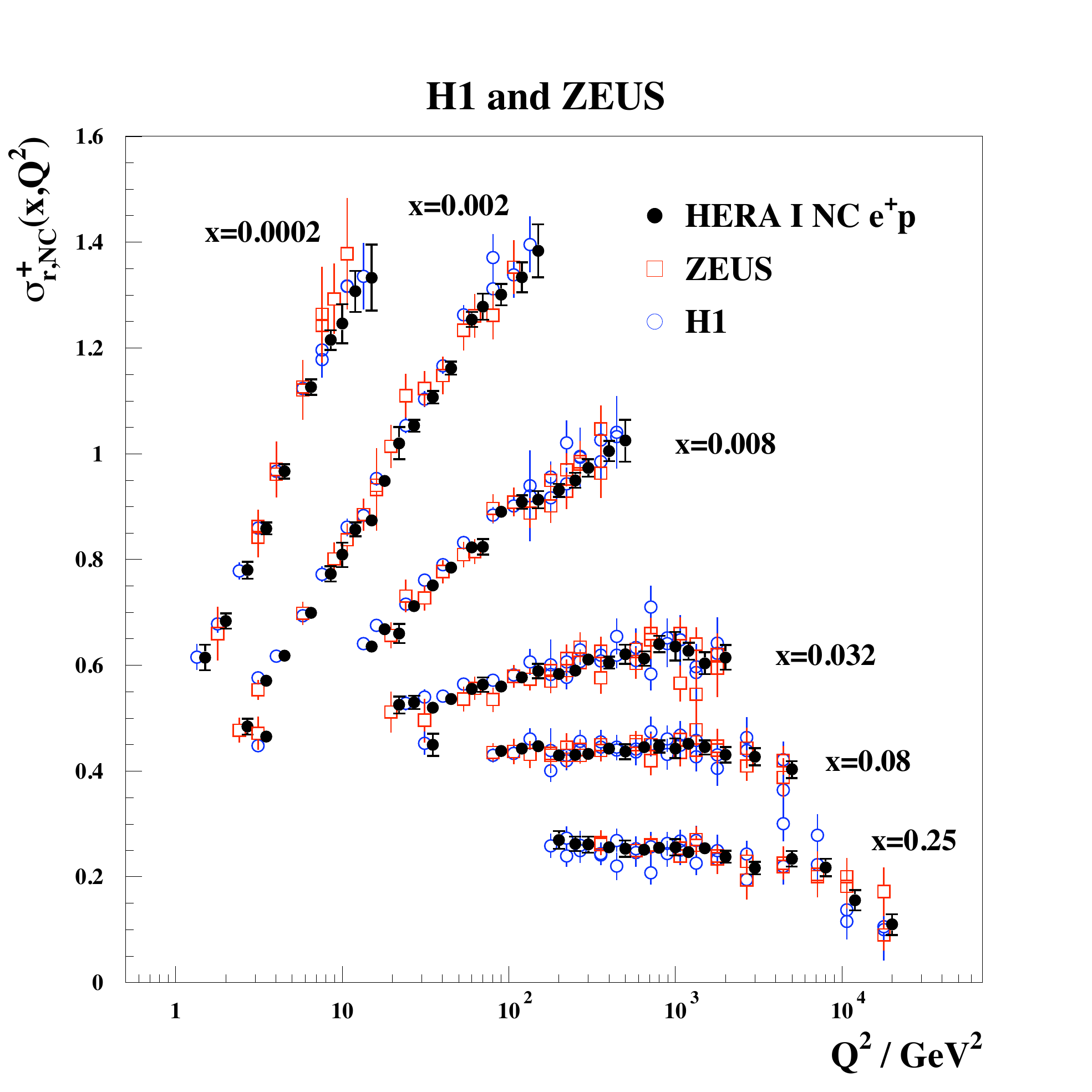}
  \caption{\label{fig:physcase} Examples illustrating the physics case for the preservation of
    HEP data. (a) Recent measurements of the strong coupling, $\alpha _{s}$ from an event shape
    analysis of JADE data at various centre of mass energies, $\sqrt{s}$. The full and dashed
    lines indicate the result from the JADE NNLO analysis. The results from a recent NNLO analysis
    of ALEPH data are also shown. (b) Observed and expected limits from ALEPH on the combined
    production cross section times branching ratio in the search for the process
    $h \rightarrow  2a \rightarrow 4\tau$, as a function of Higgs boson mass, $m_{h}$.
    (c) The energy dependence of the average charged-hadron $P_{T}$, featuring data from multiple
    experiments including CMS data at several centre of mass energies from the first years of LHC running.
    (d) Neutral current $e^+p$ reduced cross sections as a function of $Q^2$ for different $x$ bins.
    The measurements are made using combined H1 and ZEUS data from HERA.}
 \begin{picture} (0.,0.)
  \setlength{\unitlength}{1.0cm}
    \put (-4.5,13.5){\bf\normalsize (a)}
    \put (5.7,13.5){\bf\normalsize (b)}
    \put (-2.0,7.0){\bf\normalsize (c)}
    \put (2,6.5){\bf\normalsize (d)}
 \end{picture}
 \vspace{-1cm}
\end{figure}
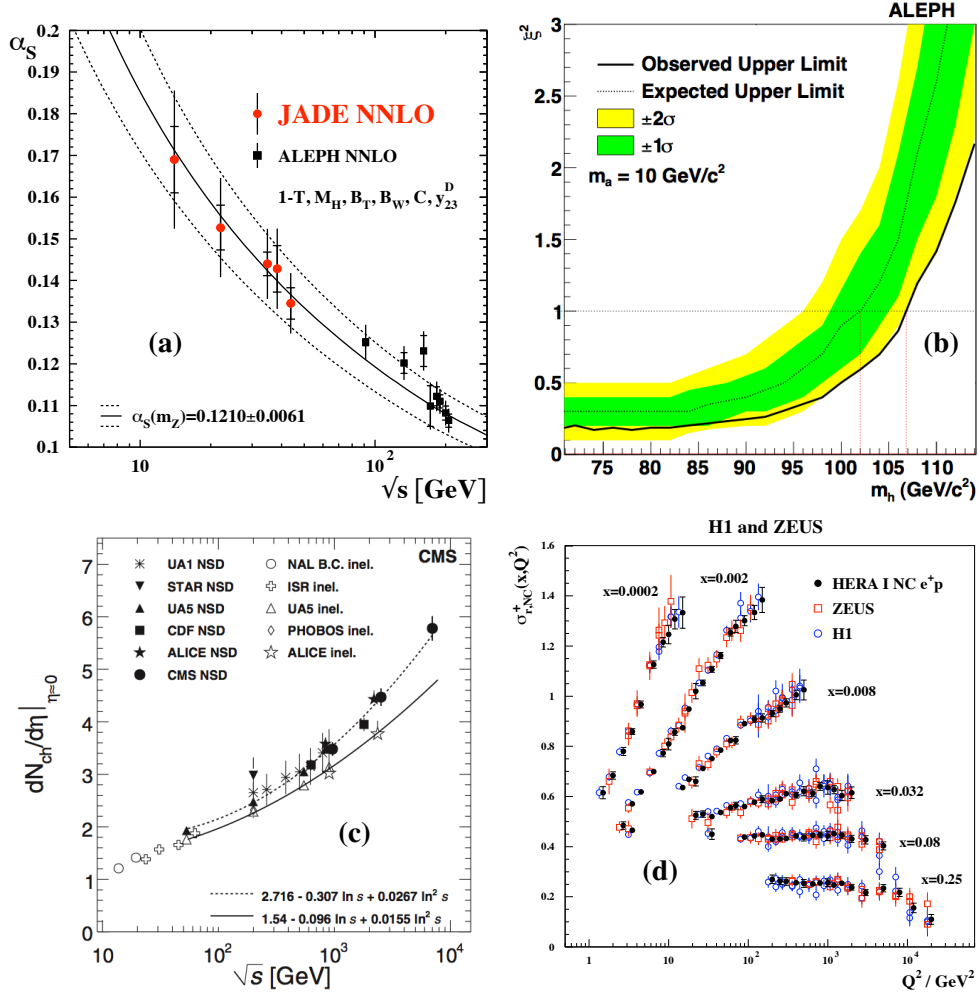

It is often assumed that older HEP data will always be superseded by that from the next generation experiment.
However, unique data sets are available in terms of initial state particles or centre of mass energy or both, such as
those from PETRA ($e^+e^-$, data taking ended in 1986), HERA and the Tevatron, as well as data from a variety of
fixed target experiments.
It may be desirable to revisit old measurements or perform new ones with such data, to achieve an increased precision
via new and improved theoretical calculations (MC models) or to explore newly developed analysis techniques.
A re-analysis of the JADE data taken at PETRA has lead to a significant improvement in the
determination of the strong coupling $\alpha_{s}(M_{Z})$, as shown in figure~\ref{fig:physcase}(a), in an
energy range that is still unique~\cite{jade,jade2}.
The running of the strong coupling, demonstrating the concept of asymptotic freedom~\cite{asym,asym2} and
in agreement with the QCD prediction, is visible from the JADE data alone: something which was not possible at
the time of the original analysis.
Results from a similar analysis by the ALEPH experiment on LEP are also shown~\cite{alephalphas}.
In a situation that mirrors the JADE analysis, it is hoped that the uncertainty on $\alpha_{s}$ will be
further reduced at some point in the future by re-analysing the very accurate HERA data once improved
theoretical predictions become available.


A further example from ALEPH is the search for the production and non-standard decay of a
Higgs boson~\cite{alephhiggs}.
A possible four tau final state is investigated, resulting from the decays of two intermediate pseudoscalars
produced via a next-to-minimal supersymmetric Standard Model Higgs decay~\cite{nmssm,nmssm2}.
For a pseudoscalar mass $m_{a} = 10$~GeV, Higgs masses $m_{h} < 107$~GeV are excluded at $95\%$
confidence level, as illustrated in figure~\ref{fig:physcase}(b).


More recently, early data taking at the LHC has resulted in several $pp$ collision datasets at unique centre of
mass energies such as $900$~GeV and $2.36$~TeV.
The first $7$~TeV data taken in 2010 also provide unique opportunities due to the very low pile-up conditions
compared to later data taking periods. 
Measurements using this data have been performed, such as the analysis of charged hadrons by
CMS~\cite{Khachatryan:2010us} shown in figure~\ref{fig:physcase}(c), but future analysis maybe
difficult if the data are not sufficiently prepared.


Cross-collaboration and combination of data from multiple experiments may provide new scientific results, with
improved precision and increased sensitivity.
This may occur during the active lifetime of similar experiments at one facility, such as
those at LEP, HERA, or the Tevatron, but may also occur later across larger boundaries,
such as combinations of Belle and BaBar or Tevatron and LHC data.
Figure~\ref{fig:physcase}(d) shows individual and combined H1 and ZEUS measurements of the reduced
neutral current cross section, where the improvement in the experimental uncertainties is clearly visible~\cite{f2_hera}.

The preservation of HEP data may facilitate the comparison of complementary physics results as well as
allowing the independent verification of experimental observations.
This is important if new phenomena are found in data recorded at the LHC or some other
future collider, when it may be useful or even mandatory to verify such results using older data.

Finally, the value of using real HEP data for scientific training, education and outreach
cannot be understated.
Providing a wide variety of HEP data sets for such analysis, with a corresponding wide
variety of associated exercises and teaching programmes, is one of the projects
identified by DPHEP to be implemented in the new collaboration phase
(see section~\ref{sec:dphepcollaboration}).

\section{Preservation models}
\label{sec:models}

In developing a series of preservation models, an all encompassing definition of ``HEP data'' is required.
Clearly the digital information, that is the data themselves, are crucial but previous attempts have confirmed
that the conservation of tapes is not equivalent to data preservation - although this may be the simplest part.
The range in data volume to be preserved is often a result not only of different sized
data sets, but different types of data: from the basic level raw data, through reconstructed
data, up to the analysis level ntuples.
However, providing not only the hardware to access the data but also the software and
environment to understand the data are the necessary and more challenging aspects.
If the experimental software is not available the possibility to study new observables or to
incorporate new reconstruction algorithms, detector simulations or event generators is lost.
Without a well defined and understood software environment the scientific potential of the data
may be limited.
Just as important are the various types of documentation, covering all facets of an experiment.
This includes the scientific publications in journals and online databases such as
INSPIRE~\cite{inspirenet} and arXiv, published theses, as well as a
myriad of internal documentation in manuals, internal notes, slides, wikis, news-groups and so on.

\begin{figure}[t]
  \begin{center}
    \includegraphics[width=\textwidth]{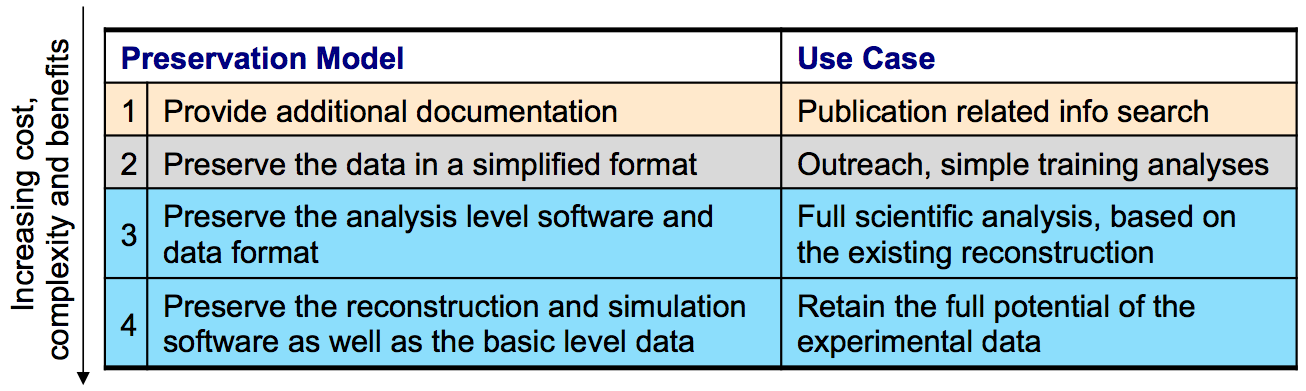}
  \end{center}
  \vspace{-0.4cm}
  \caption{Data preservation levels as defined by the DPHEP Study Group.}
 \vspace{-0.2cm}
 \label{fig:levels}
\end{figure}

Considering this inclusive definition of HEP data, a series of data preservation levels has been established
by the DPHEP Study Group, as summarised in figure~\ref{fig:levels}.
The levels are organised in order of increasing benefit, which comes with increasing complexity and cost.
Each level is associated with use cases, and the preservation model adopted by an experiment should 
reflect the level of analysis expected to be available in the future.
The four levels represent three different areas, which represent complementary initiatives:
documentation (level 1), outreach and simplified formats for data exchange (level 2)
and technical preservation projects (levels 3 and 4). 


Whereas most collaborations involved in DPHEP pursue some form of level 1 and 2 strategies,
levels 3 and 4 are really the main focus of the data preservation effort: to maintain usable access to
analysis level data, MC and the analysis level software, in addition (in the case of level 4) to the
reconstruction and simulation software.
This may be realised using two alternative paradigms: either keep the current environment alive as
long as possible or adapt and validate the code against future changes as they happen.
These two complementary approaches are taken by BaBar at SLAC and the HERA experiments at DESY,
both employing virtualisation techniques, but in different ways, as described in detail in
the 2012 DPHEP publication~\cite{dpheppub2}.
Other HEP experiments, including those at the LHC, are now examining such solutions and developing
their own long term plans for data preservation, before the collisions have stopped.

\section{Towards the DPHEP Collaboration}
\label{sec:dphepcollaboration}

The DPHEP Study Group will move to a new organisational model, the DPHEP Collaboration, in 2013 and
and the formal signing procedure of the Collaboration Agreement has now commenced.
In addition to the DPHEP Chair, a new Project Manager position has now been established,
which is initially based at CERN.
DPHEP will continue to investigate and take action in areas of coordination, preservation
standards and technologies, as well as expanding the experimental reach and
inter-disciplinary cooperation.
Often working together with other scientific disciplines or other national data preservation
programmes, areas of interest include: tools and best practices for the ingestion of (meta-)data;
making data discoverable for clearly identified communities under defined (open) access policies;
strategies and best practices for archival management.
During the next period, full deployment of the various experiment and laboratory based projects
is foreseen, including generic validation frameworks and long term storage solutions.
A multi-layered, multi-experiment DPHEP portal is also planned as a convenient interface to open
access initiatives within the HEP community.



\begin{thebibliography}{99}

\bibitem{dphep}
DPHEP, Study group on data preservation and long term analysis in HEP; {\tt http://dphep.org}.

\bibitem{dpheppub1}
 D.~Asner {\it et al.} [DPHEP Study Group],
 arXiv:0912.0255.

\bibitem{dpheppub2}
 Z.~Akopov {\it et al.} [DPHEP Study Group],
 arXiv:1205.4667.

\bibitem{jade}
S.~Bethke {\it et al}. [JADE Collaboration],
Eur.\ Phys.\  J.\ C {\bf 60} (2009) 181;\\
Erratum-ibid. C {\bf 62} (2009) 451 [arXiv:0810.2933].

\bibitem{jade2}
S.~Bethke {\it et al}. [JADE Collaboration],
Eur.\ Phys.\  J.\ C\ {\bf 64} (2009) 351 [arXiv:0810.1389].

\bibitem{asym}	
D.~J.~Gross and F.~Wilczek,
Phys.\ Rev.\ Lett.\ {\bf30} (1973) 1343.

\bibitem{asym2}
H.~D.~Politzer,
Phys.\ Rev.\ Lett.\ {\bf 30} (1973) 1346.

\bibitem{alephalphas}
G.~Dissertori {\it et al.},
 J.\ High\ Energy\ Phys.\ {\bf 0802} (2008) 040 [arXiv:0712.0327].

\bibitem{alephhiggs}
S.~Schael {\it et al.} [ALEPH Collaboration],
J.\ High\ Energy\ Phys.\ {\bf 1005} (2010) 049 [arXiv:1003.0705].

\bibitem{nmssm}
R.~Dermisek and J.~F.~Gunion,
Phys.\ Rev.\ Lett.\ {\bf 95} (2005) 041801 [hep-ph/0502105].

\bibitem{nmssm2}
R.~Dermisek and J.~F.~Gunion,
Phys.\ Rev.\ D\ {\bf 76} (2007) 095006 [arXiv:0705.4387].

\bibitem{Khachatryan:2010us}
V.~Khachatryan {\it et al.} [CMS Collaboration],
Phys.\ Rev.\ Lett.\  {\bf 105} (2010) 022002 [arXiv:1005.3299].

\bibitem{f2_hera}
F.~D.~Aaron {\it et al.} [H1 and ZEUS Collaborations],
J.\ High\ Energy\ Phys.\ {\bf 1001} (2010) 109 [arXiv:0911.0884].

\bibitem{inspirenet}
INSPIRE, High energy physics literature database; {\tt http://inspirehep.net}.

\end{thebibliography}
\end{document}